\documentclass[11pt]{article}
\usepackage{amsmath,amstext,amsbsy,amssymb}
\usepackage{bm}
\newcommand{\Ip}{I_+}
\newcommand{\Ie}{I_-}
\newcommand{\tr}{{\rm tr}}

\newcommand{\vF}{v_{\scriptscriptstyle{ F}}}
\newcommand{\td}{{\scriptscriptstyle{1+1}}}
\newcommand{\thd}{{\scriptscriptstyle{2+1}}}
\newcommand{\fd}{{\scriptscriptstyle{3+1}}}
\newcommand{\fid}{{\scriptscriptstyle{4+1}}}

\newcommand{\ssH}{{\scriptscriptstyle{H}}}
\newcommand{\ssSH}{{\scriptscriptstyle{SH}}}

\newcommand{\ssF}{{\scriptscriptstyle{F}}}

\newcommand{\ssB}{{\scriptscriptstyle{\rm B}}}

\newcommand{\ssU}{{\scriptscriptstyle{U}}}
\newcommand{\ssD}{{\scriptscriptstyle{D}}}
\newcommand{\ssT}{{\scriptscriptstyle{T}}}

\long\def\symbolfootnote[#1]#2{\begingroup%
\def\thefootnote{\fnsymbol{footnote}}\footnote[#1]{#2}\endgroup}

\begin{document}

\begin{center}

{\Large \bf Effective field theory of a topological insulator and the Foldy-Wouthuysen transformation }

\vspace{2cm}

{\bf\"{O}mer F. Dayi},
\symbolfootnote[0]{{\it E-mail addresses:} dayi@itu.edu.tr , elbistan@itu.edu.tr, yunt@itu.edu.tr }  {\bf Mahmut Elbistan},\, {\bf Elif Yunt}\\
\vspace{5mm}

{\em {\it Physics Engineering Department, Faculty of Science and
Letters, Istanbul Technical University,\\
TR-34469, Maslak--Istanbul, Turkey} }

\vspace{3cm}

{\bf Abstract}

\end{center}

\noindent 
Employing  the Foldy-Wouthuysen transformation it is  demonstrated straightforwardly that
 the first and second  Chern numbers  are equal to
the coefficients of the $2+1$ and $4+1$ dimensional Chern-Simons actions which are generated by  the massive Dirac fermions coupled to the Abelian gauge fields.
A   topological insulator model in $2+1$ dimensions is discussed and by means 
of  a dimensional reduction approach 
the $1+1$ dimensional descendant of the  $2+1$ dimensional Chern-Simons  theory  is presented.
Field strength of the Berry gauge field  corresponding to the $4+1$ dimensional Dirac theory 
is explicitly derived  through the Foldy-Wouthuysen transformation. Acquainted with it
the second Chern numbers are calculated for specific choices of the integration domain. 
A method is proposed to obtain $3+1$  and  $2+1$ dimensional descendants of the effective field theory of the
$4+1$ dimensional time reversal invariant topological insulator theory.
Inspired by the spin Hall effect in graphene, a hypothetical model of the time reversal invariant spin Hall insulator in $3+1$ dimensions  is proposed.

\vspace{1cm}

\renewcommand{\theequation}{\thesection.\arabic{equation}}
\newpage

\section{Introduction}

$2+1$ dimensional Dirac theory which was supposed to be  physically unrealizable turned out  to be indispensable to reveal the main features  of graphene.
At the Dirac points, in the low energy and long wavelength limit, the electrons of graphene effectively satisfy the  $2+1$ dimensional    Dirac equation where the velocity of light $c$ is substituted by  the effective velocity $v_\ssF$ \cite{sem, vm}. This had enormous impact on  condensed matter physics. Based on the $2+1$ dimensional    Dirac Hamiltonian Haldane\cite{hal1}  discovered that it is possible to construct a model where an integer quantum Hall effect results without an external magnetic field. In spite of the fact that a magnetic field is not needed, it is a time reversal breaking (TRB) theory. Kane and Mele\cite{km} incorporated the spin degrees of freedom into the Haldane's construction and formulated the time reversal invariant (TRI) spin Hall effect in graphene. This model\cite{km} paved the way to the theoretical prediction of the     topological insulator phase in real materials\cite{bhz} which was observed for the first time in \cite{kon}. Topological  insulators are usually defined to be ordinary insulators in the bulk which possess conducting states moving at the edge or boundary surface\cite{hk,qz,hm}. They can be classified by a new topological invariant called $\mathbb{Z}_2$\cite{kmZ2}.

Topological invariants emerge already in the quantum Hall effect. The Hall conductivity is given by the first Chern number\cite{tknn, ass}.
 Moreover, in  $2+1$ dimensions a topological gauge theory is generated
 by integrating out the massive Dirac fermion  fields coupled to Abelian gauge fields in the related path integral\cite{ns,r1,r2}. It
 is described by the $2+1$ dimensional Chern-Simons  action whose
coefficient is the winding number of the  noninteracting massive fermion propagator which
is equal to the first Chern number resulting from the Berry gauge field\cite{berry,xcn} of the quantum Hall states. 
Thus the Hall current can be acquired from a topological field theory which manifestly violates time reversal symmetry\cite{zhk,revz}. 
One can also derive the TRI spin Hall current of the model introduced in \cite{km} by calculating the related first Chern numbers\cite{oe}. 

The $4+1$ dimensional  Chern-Simons action which can be generated by the massive fermions coupled to Abelian gauge fields, is  manifestly TRI and 
Qi-Hughes-Zhang\cite{qhz} designated it as the effective topological field theory  of the fundamental TRI topological insulator in $4+1$ dimensions.
They demonstrated that for the band insulators which can be deformed adiabatically to a flat band model, the coefficient of the effective  action is equal to the second Chern number given by the related non-Abelian Berry vector fields. 
We will prove the equivalence of the coefficients of the induced Chern-Simons actions with the Chern numbers in a straightforward manner
by employing the Foldy-Wouthuysen transformation.
Qi-Hughes-Zhang constructed the $3+1$  and $2+1$ dimensional
 descendant theories  by dimensional reduction from the $4+1$ dimensional  action of the massive Dirac fields coupled to external gauge fields. We mainly follow  the approach of \cite{qhz}, though  we only deal with the continuous Dirac theory and 
propose a slightly modified method to introduce the descendant theories which permits us to derive explicitly the related physical objects like polarizations.  Moreover, we posit a hypothetical model of TRI  spin Hall effect in $3+1$ dimensions which may be useful to understand some aspects of physically realizable 
three dimensional topological insulators\cite{3fkm,3mb,3r}.

In the next section, we introduce the  Berry gauge fields corresponding to  Dirac fermions   through the   Foldy-Wouthuysen transformation.
Section \ref{topCS} is devoted to demonstrate in a straightforward fashion that  the coefficients of the 
the Chern-Simons actions generated by integrating out the massive Dirac spinor fields which are coupled to the Abelian gauge fields  in the path integral 
are equal to  the first and second Chern numbers given, respectively, by the Berry gauge fields in $2+1$  and  $4+1$  dimensions.
We consider the $2+1$ dimensional    topological field theory of the integer quantum Hall effect in Section \ref{a2dcms} and recall how to construct the TRI spin quantum Hall effect in graphene which is a model of $2+1$ dimensional topological insulator. Then, we present the dimensional reduction to $1+1$ dimensions by obtaining the
one dimensional charge polarization explicitly. 
In Section \ref{foap}, we consider the  $4+1$ dimensional   Chern-Simons field theory  which was shown to describe the fundamental topological insulator.
We derive the field strengths of the related Berry gauge fields to obtain the second Chern number and study dimensional reduction to $3+1$ dimensions.
By imitating the approach of \cite{km} we
theorized a hypothetical model in $4+1$ dimensions  which yields a TRI spin Hall current in $3+1$ dimensions
by means of  the dimensional reduction. Slightly modifying the approach of \cite{qhz} we proposed a dimensional reduction procedure to $2+1$ dimensions
which provides explicit forms of the gauge field components  which take part in the descendant action.
In the last section we discussed the results obtained.

\section{The Foldy-Wouthuysen transformation and the Berry gauge field \label{foldyw}}

We consider relativistic electrons of charge $e>0$ with a characteristic velocity  like the velocity of light $c$ or 
the effective velocity $\vF$ as in graphene. To retain the formulation general we work in units $\hbar=c=\vF=1,$ as well as $e=1$ and recuperate them when needed.
Thus, the free, massive electrons are described by the Dirac Hamiltonian 
\begin{equation}
  H= \bm{\alpha}\cdot \bm{k}+ \beta m .
 \label{DH}
 \end{equation}
In this section, vectors  are $d$-dimensional, like the momentum $\bm k$ whose components are  denoted  $k_I;\ I=1,\cdots , d.$ 
The Hamiltonian (\ref{DH}) can be diagonalized as
\begin{equation}
\label{diag}
 UHU^\dagger =E\beta,
\end{equation}
where $E$ is the  total energy
\begin{equation}
\label{energy}
E= \sqrt{k^2+m^2 },
\end{equation}
and $U$ is the unitary Foldy-Wouthuysen transformation
$$
U=\frac{\beta H+E}{\sqrt{2E(E+m)}} .
$$
Through  the  transformation $U$ a pure gauge field can be introduced as
\begin{equation}
\label{GF}
\bm{{\cal A}}^\ssU=i U(\bm{k})\frac{\partial U^\dag(\bm{k})}{\partial \bm{k}}.
\end{equation}
The  Berry gauge field $\bm{{\cal A}}$ follows by projecting (\ref{GF})  onto the positive  energy eigenstates of the Dirac Hamiltonian (\ref{DH}).
One can be convinced that eliminating the negative energy states 
is equivalent to the adiabatic approximation 
by revoking its
similarity  to   suppression of the interband interactions  in  molecular problems\cite{bm}. Thus, we define the Berry  gauge field as  
\begin{equation}
\label{BGF}
   \bm{{\cal A}} \equiv \Ip \bm{{\cal A}}^\ssU \Ip ,
\end{equation}
where $\Ip$ is the projection operator onto the positive energy subspace.
This definition of the Berry gauge field is valid irrespective of the dimensions of the Hamiltonian (\ref{DH}). In order to derive 
$\bm{{\cal A}}$ explicitly let us adopt the following $2^N\times2^N;\ N=[\frac{d}{2}],$ dimensional realizations of $\bm{\alpha}$ and $\beta$ 
\begin{align}
\label{alphabeta}
\bm\alpha=\left(
\begin{array}{cc}
  0 & \bm\rho \\
 \bm \rho ^\dagger & 0 \\
\end{array}
\right),\ \ \
\beta=
\left(
\begin{array}{cc}
   1 & 0 \\
  0 & - 1 \\
\end{array}
\right).
\end{align} 
Here  $\bm\rho$ and the unit matrix $ 1$ are $2^{N-1}\times2^{N-1}$ dimensional. 
In representation (\ref{alphabeta}) the gauge field (\ref{GF})  becomes
\begin{equation}
\label{GFgeneral}
{\cal A}^U_I=\frac{i }{2E^{2}(E+m)}\left[ E(E+m)\alpha_I\beta+ \beta
    \bm\alpha \cdot \bm k k_I   
    -iE\sigma_{IJ}k_J \right] ,
\end{equation}
where $\sigma_{IJ}\equiv -\frac{i}{2}[\alpha_I,\alpha_J]$.
Therefore, the Berry gauge field (\ref{BGF}) results to be
\begin{equation}
\label{BGFgeneral}
{\cal A}_{I} = -\frac{i}{4E(E+m)}(\rho_I \rho^\dagger_J-\rho_J \rho^\dagger_I)k_J.
\end{equation}
Although the field strength of (\ref{GFgeneral}) vanishes because of being a pure gauge field, the Berry curvature 
\begin{equation}
\label{fij}
{\cal F}_{IJ} =\frac{\partial {\cal A}_J}{\partial k_I}- \frac{\partial {\cal A}_I}{\partial k_J}-i[{\cal A}_I,{\cal A}_J] ,
\end{equation}
  is  non-vanishing in general. 

When we deal with  $2n+1$ dimensional space-time coordinates where $n=1,2\cdots ,$ the Berry curvature (\ref{fij})  can be employed to define the  Chern number
which is the integrated Chern character, as\cite{nak}  
\begin{equation}
\label{NG}
N_n=\frac{1}{(4\pi)^n n!} 
\int_{{\cal M}_{2n}} d^{2n}k\ \epsilon_{I_1I_2 \cdots  I_{2n}}\tr \left\{ {\cal F}_{I_1I_2}\cdots {\cal F}_{I_{2n-1}I_{2n}} \right\}.
\end{equation}
 For  $2+1$ dimensional    systems  the Berry gauge field is Abelian, so that ${\cal F}_{ab}=\partial {\cal A}_b/ \partial k_a -\partial {\cal A}_a/ \partial k_b,$
 where $a,b=1,2,$ and the first Chern number is
\begin{equation}
\label{C1def}
N_1=\frac{1}{4\pi}\int{d^2k\epsilon_{ab}\tr {\cal F}_{ab}}.
\end{equation}
In $4+1$ dimensions   one introduces the  second Chern number  as
\begin{equation}
\label{C2def}
N_2=\frac{1}{32\pi^2}\int{d^4k\epsilon_{ijkl}\tr\{{{\cal F}_{ij}{\cal F}_{kl}}\}},
\end{equation}     
where $i,j,k,l=1,2,3,4.$

\setcounter{equation}{0}
\section{Topological field theories and the Chern numbers \label{topCS}}

Field theory of electrons  interacting with the external Abelian gauge field $A_\alpha$ 
is given  by the  Dirac Lagrangian density
\begin{equation}
{\cal L} \left(\psi,\bar{\psi}, A \right) =\bar{\psi} \left[ \gamma^\alpha \left( p_\alpha + A_\alpha \right)- m \right] \psi ,
\label{lag}
\end{equation}
where $\alpha =0,1\cdots d.$ 
By integrating out the fermionic degrees of freedom 
in the related path integral one formally gets the action of the external fields as
\begin {equation}
S[A]=-i\ln \det[i\gamma^\alpha(\partial_\alpha-iA_\alpha)-m].
\end{equation} 
For $d=2n$ one of the terms which  it gives rise to is  
$$
T[A^{\scriptscriptstyle{n+1}}]=\int [dq_1]\cdots [dq_{n+1}] A^{\alpha_1}(q_1)\cdots A^{\alpha_{n+1}}(q_{n+1})\pi_{\alpha_1\cdots  \alpha_{n+1}}(q_1\cdots q_{n+1}).
$$
$[dq]$ denotes the integral over the related phase space. At the order of first loop 
$$
\pi_{\alpha_1\cdots  \alpha_{n+1}}(q_1\cdots q_{n+1})=\int {\frac{d^{2n+1}k}{(2\pi)^{2n+1}}}\tr\{G(k)\lambda_{\alpha_1}(k,k-q_1)G(k-q_1)\cdots \lambda_{\alpha_{n+1}}(k+q_{n+1},k)\} ,
$$
where $G(k)$ is the one particle Green function of the free Dirac field and $\lambda_{\alpha}$ is the photon vertex.
$T[A^{\scriptscriptstyle{n+1}}]$ generates  the $2n+1$ dimensional Chern-Simons term
\begin{equation}
\label{CS}
S{_{eff}^{\scriptscriptstyle{2n+1}}}[A]=C_{n}\int d^{2n+1}x\epsilon^{\alpha_1\cdots\alpha_{2n+1}}A_{\alpha_1}\partial_{\alpha_2}A_{\alpha_3}\cdots \partial_{\alpha_{2n}}A_{\alpha_{2n+1}},
\end{equation}
which can be taken as the effective topological action in the low energy limit.
In the weak field approximation the coefficient $C_n$ can be written as \cite{gjk}
$$
C_n=\frac{\epsilon^{\alpha_1\beta_1\cdots \alpha_n\beta_n\alpha_{n+1}}}{(n+1)(2n+1)!}\partial_{(1)\beta_1}\cdots\partial_{(n)\beta_n}\pi_{\alpha_1\cdots  \alpha_{n+1}}(q_1\cdots q_{n+1})|_{q_i=0}    ,
$$
where $\partial_{(n)\alpha}\equiv \partial /\partial q^\alpha_n .$ 
Substituting the photon vertex with
$$
\lambda_{\alpha}(k,k)=-i\partial_\alpha G^{-1}(k) ,
$$
where $G^{-1} (k)$ is the inverse of $G(k)$, one can express $C_n$ as the winding number of the fermion propagator $G$ \cite{gjk}:
\begin{equation} 
\label {CN}
C_{n}=\frac{(-i)^{n+1}\epsilon^{\alpha_1...\alpha_{2n+1}}}{(n+1)(2n+1)!}\int{\frac{d^{2n+1}k}{(2\pi)^{2n+1}}\tr\{[G(k)\partial_{\alpha_1}G(k)^{-1}]...[G(k)\partial_{\alpha_{2n+1}}G(k)^{-1}]\}} .
\end{equation}

In the rest of this section we will demonstrate that for the $2+1$  and $4+1$ dimensional     Dirac theories
the topological invariant winding numbers of the fermion propagator  (\ref{CN}) are equal to the Chern numbers of the Berry gauge fields (\ref{NG}),
up to constant factors. Obviously, these results can be obtained by other means; however we think that the following derivations are quite straightforward 
and clear.

By virtue of the Foldy-Wouthuysen transformation $U,$ one can invert   
(\ref{diag}) to write  the Dirac Hamiltonian  as $H=EU^\dagger \beta U,$ which is suitable to
express it in a projector form. Indeed, for $k^{\alpha }=(w,\bm k)$ we can write the inverse of the propagator as 
\begin{equation}
\label{hpr}
G^{-1}(k)=w+(E+i\varepsilon )(P_--P_+),
\end{equation}  
where the projection operators $P_+$ and  $P_-,$  satisfying
\begin{equation}
\label{proj}
P_+^2=P_+,\ P_-^2=P_-,\ P_++P_-= 1,
\end{equation}
are given explicitly as
\begin{equation}
\label{pape}
P_+
=U^{\dagger}\Ip  U,\ \ 
P_-
=U^{\dagger}\Ie  U .
\end{equation}
The  operators $\Ip  $ and $\Ie  $ project, respectively, onto the positive and the negative energy states which are obtained
in  representation (\ref{alphabeta}) as
\begin{eqnarray*}
\Ip  
=\left(
\begin{array}{cc}
  1 & 0 \\
  0 & 0 \\
\end{array}
\right), 
&
\Ie  
=\left(
\begin{array}{cc}
  0 & 0 \\
  0 & 1 \\ 
\end{array}
\right) .
\end{eqnarray*}
Now, (\ref{hpr}) can easily be inverted to obtain the Green function as
$$
G(k)={\frac{P_+}{w-(E+i\varepsilon )}}+{\frac{P_-}{w+(E+i\varepsilon )}}.
$$
In the sequel  we will not explicitly write the positive, 
infinitesimal parameter $\varepsilon $ unless necessary.  

Derivative of $G^{-1}$ with respect to $k_0=w$ is 
\begin{equation}
\frac{\partial G^{-1}(k)}{\partial w}=1 .
 \label{pm1} 
\end{equation}
Moreover, owing to the projector form (\ref{hpr}) and  the energy-momentum relation (\ref{energy}), it  satisfies
\begin{equation}
{\frac{\partial G^{-1}(k)}{\partial k_I}}=\frac {k_I}{E}(P_--P_+)- 2E\frac{\partial P_+}{\partial k_I}. \label{pm}
\end{equation}
The following relations between the projection operators $P_+$ and $P_-$ 
$$
P_+\frac{\partial P_-}{\partial k_I}=-\frac{\partial P_+}{\partial k_I}P_-=\frac{\partial P_-}{\partial k_I}P_-;\ \ \
P_-\frac{\partial P_-}{\partial k_I}=-P_-\frac{\partial P_+}{\partial k_I}=\frac{\partial P_-}{\partial k_I}P_+ ,
$$
can easily be derived by inspecting their basic properties (\ref{proj}).

\subsection{Relation between  $C_1$ and $N_1$}

In $2+1$ dimensions   integration of the massive Dirac fermions in the related path integral with the  Lagrangian density  (\ref{lag}) leads to
the effective topological action
\begin{equation}
\label{Ef2}
S{_{eff}^{\thd}}[A]=C_1\int d^3x \epsilon^{\mu\nu\rho}A_\mu\partial_\nu A_\rho ,
\end{equation}
where  $\mu ,\nu, \rho =0,1,2.$     
The coefficient $C_1$  is given by
\begin{equation}
\label{cbr}
C_{1}=-\frac{1}{12}\epsilon^{\mu\nu\rho}\int \frac{d^2kdw}{(2\pi)^3}\tr\{[G(k)\partial_{\mu}G(k)^{-1}][G(k)\partial_{\nu}G(k)^{-1}][G(k)\partial_{\rho}G(k)^{-1}]\} .
\end{equation}
Making use of (\ref{pm1}) we can express (\ref{cbr}) as 
$$
C_{1}=-\frac{1}{4}\epsilon_{ab}\int{\frac{d^2kdw}{(2\pi)^3}\tr\{G(k)G(k)\partial_aG(k)^{-1}G(k)\partial_bG(k)^{-1}\}}.
$$
When we write the integrand explicitly by employing (\ref{pm}), obviously the quadratic terms in $k_i$ vanish, so that we get
\begin{eqnarray*}
C_1&=&-\frac{1}{4}\epsilon_{ab}\int\frac{d^2kdw}{(2\pi)^3} \tr\{(\frac{P_+}{(w-E)^2}+\frac{P_-}{(w+E)^2}
    )(4E^2\partial_a P_+G\partial_b P_+ \nonumber \\  
    &&\nonumber \\  
    && -2k_a[P_--P_+]G\partial_b P_+ -2k_b\partial_a P_+G[P_--P_+])\} .
\end{eqnarray*}
One can observe that $k_a$ and $k_b$ terms   combine to vanish
$$
2\epsilon_{ab}\tr\{[\frac{P_+}{(w-E)^3}-\frac{P_-}{(w+E)^3}][k_a\partial_b P_++k_b\partial_aP_+]\}=0,
$$
even before performing the  $w$ integration.
The remaining terms, after revoking the infinitesimal parameter $\varepsilon,$ become
$$
C_{1}=-\epsilon_{ab}\int{\frac{d^2kdw}{(2\pi)^3}E^2\tr\{(\frac{P_+}{(w-E-i\varepsilon )^2(w+E+i\varepsilon )}+\frac{P_-}{(w-E-i\varepsilon )(w+E+i\varepsilon )^2} )\partial_aP_+\partial_bP_+\}}.
$$
Integration over $w$  yields
$$
C_1=\frac{i}{8\pi^2}\epsilon_{ab}\int d^2k\tr\{P_+\partial_aP_+\partial_bP_+\}.
$$
Making use of   definitions (\ref{pape}),  one can easily observe that
\begin{eqnarray*}
\epsilon_{ab}\tr \{P_+\partial_aP_+\partial_bP_+\} 
&=& \epsilon_{ab}\tr\{(\Ip  U\partial_aU^{\dagger})(\Ip  U\partial_bU^{\dagger})
+ \Ip  \partial_aU\partial_bU^{\dagger}\Ip \}\\
&=& \epsilon_{ab}\tr\{\Ip  \partial_aU\partial_bU^{\dagger}\Ip \}.
\end{eqnarray*}
On the other hand, plugging the field strength  (\ref{fij}) of
the Abelian Berry gauge field (\ref{BGF}) into (\ref{C1def}) leads to
$$
N_1=\frac{i}{2\pi}\int d^2k\epsilon_{ab}\tr\{\Ip  \partial_aU\partial_bU^{\dagger}\Ip \}  .
$$
Therefore, we conclude  that
\begin{equation}
\label{c1n1}
C_1=\frac{N_1}{4\pi}.
\end{equation} 

\subsection{Relation between $C_2$ and $N_2$}

When we deal with $4+1$ dimensions, the effective topological action (\ref{CS}) becomes
$$
S_{eff}^{\fid}[A]=C_2\int d^5x \epsilon^{ABCDE}A_A \partial_B A_C \partial_D A_E ,
$$
where $A,B,\cdots=0,\cdots, 4.$ $C_2$ is given by (\ref{CN}) 
  as
\begin {equation}
\label{cc2}
C_2=\frac{i}{3\times5!} \int \frac{d^4kdw}{(2\pi)^5}\epsilon ^{ABCDE} \tr\{G\partial_A G^{-1}G\partial_B G^{-1}G\partial_C G^{-1}G
\partial_D G^{-1}G\partial_E G^{-1}\}.
\end{equation}
Plugging (\ref{pm1}) and
 (\ref{pm}) into (\ref{cc2})  leads to
\begin{eqnarray*}
C_2&=&\frac{i}{3\times 4!}\epsilon_{ijkl}\int\frac{d^4k dw}{(2\pi)^5} \tr\{GG[\frac{k_i}{E}(P_--P_+)-2E\partial_i P_+] G\nonumber \\
&&[\frac{k_j}{E}(P_--P_+)-2E\partial_j P_+]G
[\frac{k_k}{E}(P_--P_+)-2E\partial_k P_+] G
[\frac{k_l}{E}(P_--P_+)-2E\partial_l P_+]\} .
\end{eqnarray*} 
Because of the symmetry properties the terms depending on $\bm k$ at the second or higher 
order   vanish. Thus, we need to consider only the following terms
\begin{eqnarray*}
C_2&=&\frac{i}{9(2\pi)^5}\epsilon_{ijkl}\int d^4k dw
\tr\{2E^4GG\partial_iP_+G\partial_jP_+G\partial_kP_+G\partial_lP_+ \nonumber \\
&&-E^2\big( k_iGG[P_--P_+]G\partial_jP_+G\partial_kP_+G\partial_lP_+ 
+k_jGG\partial_iP_+G[P_--P_+]G\partial_kP_+G\partial_lP_+ \nonumber \\
&&+k_kGG\partial_iP_+G\partial_jP_+G[P_--P_+]G\partial_lP_+
+k_lGG\partial_iP_+G\partial_jP_+G\partial_kP_+G[P_--P_+]\big)\} .
\end{eqnarray*}

Inspecting the symmetry properties one can easily observe that
the second and the fifth terms are summed to give a vanishing contribution.
Similarly,  one can show that gathered together contributions of the third  and the fourth terms vanish by making use of the equalities
\begin{eqnarray*}
&\partial_iP_+G(P_--P_+)=\left(\frac{P_+}{w+E}-\frac{P_-}{w-E}\right)\partial_iP_+,\ \ \   
(P_--P_+)G\partial_lP_+=\partial_lP_+\left(\frac{P_+}{w+E}-\frac{P_-}{w-E}\right) .&
\end{eqnarray*}
Hence, we get
$$
C_2=\frac{2i}{9(2\pi)^5}\epsilon_{ijkl}\int d^4kdw
E^4\tr[GG\partial_i P_+G\partial_j P_+G\partial_k P_+G\partial_lP_+].
$$
After restoring  $\varepsilon $  this can be expressed as
\begin{eqnarray*}
&C_2  =  \frac{2i }{9(2\pi)^5}\epsilon_{ijkl}\int d^4kdw 
E^4\tr[(\frac{P_+}{(w-E-i\varepsilon)^3(w+E+i\varepsilon)^2}+\frac{P_-}{(w-E-i\varepsilon)^2(w+E+i\varepsilon)^3}) 
\partial_i P_+\partial_j P_+\partial_k P_+\partial_lP_+]. &
\end{eqnarray*}    
By performing the $w$ integration we obtain 
\begin{equation}
\label{prep1}
C_2=-\frac{1}{12}\epsilon_{ijkl}\int \frac{d^4k}{(2\pi)^4} \tr\{\partial_iP_+P_-\partial_jP_+\partial_kP_+P_-\partial_lP_+\}.
\end{equation}
In terms of the explicit forms of $P_+$ and $P_-$ given by (\ref{pape})  we  can express the integrand of (\ref{prep1}) as
\begin{eqnarray}
\epsilon_{ijkl}\tr\{\partial_iP_+P_-\partial_jP_+\partial_kP_+P_-\partial_lP_+\}&=&\epsilon_{ijkl}\tr\{(U^{\dagger}\Ip  \partial_iU+\partial_iU^{\dagger}\Ip  U)
P_-(U^{\dagger}\Ip  \partial_jU+\partial_jU^{\dagger}\Ip  U)\nonumber \\
& &(U^{\dagger}\Ip  \partial_kU+\partial_kU^{\dagger}\Ip  U)
P_-(U^{\dagger}\Ip  \partial_lU+\partial_lU^{\dagger}\Ip  U)\}\nonumber \\
&=&\epsilon_{ijkl}\tr\{\Ip  \partial_iUP_-\partial_jU^{\dagger}\Ip  \partial_kUP_-\partial_lU^{\dagger}\Ip  \}\label{prep2}     .               
\end{eqnarray}
Now, the  Berry gauge field (\ref{BGF}) is non-Abelian and its
curvature ${\cal F}_{ij}$  can be written as
\begin{eqnarray*}
{\cal F}_{ij} &= & i\Ip \partial_iU\partial_jU^\dagger\Ip+i\Ip  U\partial_iU^{\dagger}\Ip  U\partial_jU^{\dagger}\Ip   -i\leftrightarrow j\nonumber \\
 &= & i\Ip \partial_iUP_-\partial_jU^\dagger\Ip -
 i\Ip \partial_jUP_-\partial_iU^\dagger\Ip .
\end{eqnarray*} 
Inserting it into the definition of the second Chern number (\ref{C2def}) and inspecting (\ref{prep1}) and (\ref{prep2})
one concludes that 
\begin{equation}
\label{c2n}
C_2=\frac{N_2}{24\pi^2}.
\end{equation}

Generalizing this method  to  higher dimensions is   straightforward.

\setcounter{equation}{0}
\section{$\bm {2+1}$ dimensional theory and  dimensional reduction to $\bm {1+1}$ dimensions   \label{a2dcms}}

One can observe that by employing (\ref{c1n1}) in (\ref{Ef2}) the effective
topological action of external gauge fields coupled to massive Dirac electrons living in $2+1$ dimensions  becomes
\begin{equation}
\label{cs3}
S^{\thd}_{eff} =\frac{N_1}{4\pi}\int d^3x \epsilon^{\mu\nu\rho}A_\mu \partial_\nu A_\rho.
\end{equation} 
To calculate the related first Chern number   (\ref{C1def}), let us choose the representation $\bm \alpha=(\sigma_x,\sigma_y),$ where $\sigma_a$ are the Pauli spin matrices.
This corresponds to set $\rho_{a}=(1,-i)$ in (\ref{BGFgeneral}). Thus, 
the Abelian Berry gauge field can be written as
\begin{equation}
\label{BGF2+1}
 {\cal A}_{a}=\frac{\epsilon_{ab} k_b}{2E(E+m)} .
\end{equation}
It yields the Berry curvature 
\begin{equation}
\label{Fberry2+1}
{\cal F}_{12} =\left(\frac{\partial {\cal A}_2}{\partial k_1}- \frac{\partial {\cal A}_1}{\partial k_2}\right) = -\frac{m}{2E^3}  .
\end{equation}
We plug (\ref{Fberry2+1}) into (\ref{C1def}) and perform the change of variable by (\ref{energy})
to express the related first Chern number as
\begin{equation}
\label{C1M}
    N_1=-\frac{m}{2} \int_{\ssD} \frac{dE}{E^2},
\end{equation}
where the domain of integration $D$ will be specified according to the model considered. If it is required to treat the  $E>0$ and $E<0$ domains on the same footing, we  can  deal with 
\begin{equation}
\label{C1}
N_1=-\frac{m}{2} \int^{m}_{-\infty} \frac{dE}{E^2}-\frac{m}{2} \int^{\infty}_{-m} \frac{dE}{E^2}=1.
\end{equation}

\subsection{A model for $\bm{2+1}$-dimensional  topological insulator \label{shg}}

Before presenting the graphene model of \cite{km}, let us briefly recall the interconnection between the quantum Hall effect and the Chern-Simons action in $2+1$ dimensions.
For electrons moving on a surface in the presence of the external in-plane electric field $\bm E =(E_x,E_y, 0)$ and the perpendicular magnetic field 
$\bm{ {\cal B} }= (0,0,{\cal B}_z )$ the Hall current is given by
\begin{equation}
\label{hc3}
j_a=\sigma_\ssH \epsilon_{ab}E_b .
\end{equation}
Ignoring the spin of electrons the Hall conductivity is a topological invariant\cite{tknn, ass}:
\begin{equation}
\label{hc1}
\sigma_{\ssH }=\frac{e^2}{h} N_1 .
\end{equation}
Here $N_1$ is the first Chern number  resulting from  the field strength $\cal{F}_{\ssB}$ of the Berry gauge field obtained 
from the single particle Bloch wave functions which are  solutions of the Schr\"{o}dinger equation in the presence of the external magnetic field ${\cal B}_z,$ 
integrated over the  states up to the Fermi level $E_F$  as
\begin{equation}
\label{c11}
N_1=\int^{E_\ssF}\frac{d^{2}k}{(2\pi )^{2}}{\cal F}_{\ssB} .
\end{equation}
A field theoretic  description is possible in terms of the Chern-Simons action (\ref{cs3}) with  definition (\ref{c11}).
In fact, the current obtained from the topological field theory (\ref{cs3}),
$$
j_\mu =\frac{N_1}{2\pi}\epsilon_{\mu\nu\rho}\partial^\nu A^\rho,
$$
gives for $E_a =\partial_a A_0- \partial_0 A_a$ the Hall current (\ref{hc3}). It also leads to the charge density $j_0=\sigma_\ssH B ,$ where the induced magnetic field  is $B=\partial_xA_y-\partial_yA_x.$  Note that $B$ would also be generated by the Hall current (\ref{hc3})
through the  current conservation condition   $\partial_a j_a=-\partial_tj_0.$
We would like to emphasize the fact that the field theory (\ref{cs3}) is not aware of the external magnetic field ${\cal B}_z.$  External magnetic field is responsible of creating the energy spectrum  whose consequences are encoded in the calculation of the first Chern number (\ref{c11}).

By employing the Berry gauge field derived from  the Dirac equation (\ref{BGF2+1}), we can still get the Hall conductivity as in (\ref{hc1})  by
an appropriate choice of the domain of  integration $D$ in (\ref{C1M}). This construction does not necessitate an external magnetic field.
For the first time in \cite{hal1}, Haldane described  how to obtain the quantum Hall effect without a magnetic field (vanishing in the average) through a Dirac like theory. 
To calculate the Hall conductivity following from the Dirac equation we let all the negative energy levels be occupied up to the Fermi level $E_\ssF =m$
in  (\ref{C1M}), so that 
\begin{equation}
\label{C11}
    \sigma_{\ssH }= \frac{e^2}{h}\left( -\frac{m}{2} \int^{m}_{-\infty} \frac{dE}{E^2}\right)=\frac{e^2}{2h}.
\end{equation}
In \cite{km}, Kane and Mele incorporated the spin of electrons into the Haldane model\cite{hal1} and proposed the following Hamiltonian 
for graphene
\begin{equation}
\label{H2}
H_{G} =\sigma_{x}\tau_{z}k_{x}+\sigma_{y}k_{y}+m\sigma_{z}\tau_{z}s_{z},
\end{equation}
which leads to a TRI spin current.
The mass term is generated by a spin-orbit coupling.
The Pauli spin matrices $ \sigma_{x,y,z}$ act on the states of sublattices. The  matrix $ \tau_z={\rm diag} (1,-1)$ denotes 
the Dirac points $K,\ K^\prime$ which  should be interchanged under the time reversal transformation.   The other Pauli matrix $s_z={\rm diag} (1,-1)$ describes the third component of the spin of electrons which should also be inverted under  time reversal transformation.
Thus the time reversal  operator is given by $T=UK$ where we can take $U=\tau_ys_y $  and $K$ takes the complex conjugation as well as
maps $\bm k\rightarrow -\bm k.$  Therefore  (\ref{H2}) is TRI:
$$
TH_GT^{-1} =H_G.
$$
The Abelian Berry gauge field obtained from the Hamiltonian (\ref{H2}) can be written as\cite{oe}
\begin{equation}
\label{bgfi}
 {\cal A}_{a}=\frac{1}{2E(E+m)}\epsilon_{ab} k_b { 1}_\tau s_z  ,
\end{equation}
where $1_\tau$ is the unit matrix in the $\tau_z$ space. The corresponding field strength  is
\begin{equation}\label{Fberry_2+1_QSH}
{\cal F}_{12} \equiv {\rm diag} ({\cal F}^{\uparrow}_{+},
{\cal F}^{\uparrow}_- ,{\cal F}^{\downarrow}_{+} , {\cal F}^{\downarrow}_{-}) = -\frac{m}{2E^3} { 1}_\tau s_z  .
\end{equation}  
The indices $\uparrow \downarrow$ and $\pm$ label, respectively, the third component of the spin and  $\tau_z.$
The spin current  defined as
$$
\bm j^s=\bm j^{\uparrow}_++\bm j^{\uparrow}_--\bm j^{\downarrow}_{+}-\bm j^{\downarrow}_{-},
$$
leads to the spin Hall current  
$$
j^s_a=\sigma_\ssSH\epsilon_{ab}E_b.
$$
The difference of the related first Chern numbers
\begin{eqnarray}
\label{C1_QSH}
    \Delta N_1 &=&\frac{1}{2\pi} \int_{E=-\infty}^{E=m} d^{2}k
    \left[ ({\cal F}^{\uparrow}_{+}+{\cal F}^{\uparrow}_-)-({\cal F}^{\downarrow}_{+} +{\cal F}^{\downarrow}_{-} )\right] \nonumber \\
     &=& (\frac{1}{2}+\frac{1}{2})-(-\frac{1}{2}-\frac{1}{2})=2,
\end{eqnarray}
gives  the spin Hall conductivity
$\sigma_{\ssSH}$ as
\begin{equation}
\sigma_{\ssSH}=\frac{e}{4\pi}\Delta N_1=\frac{e}{2\pi} .
\label{spcg}
\end{equation}

\subsection{Dimensional reduction to $\bm {1+1}$ dimensions  \label{dr11}}

We would like to discuss dimensional reduction from $2+1$  to $1+1$ dimensions   by slightly modifying
the procedure described in \cite{qhz}.
The dimensionally reduced theory can be defined through the
 $1+1$ dimensional   Lagrangian density  
$$
{\cal L}_\td \left(\psi,\bar{\psi},A\right) =\bar{\psi} \left[\gamma^r\left(p_r+A_r\right)+\gamma_2 \zeta_y -m\right]\psi ,
$$
where $r=t,x$ and the external field $\zeta_y (t,x)$ is the reminiscent of the gauge field $A_y.$
 We define $\zeta_y=k_y+\zeta ,$
where $k_y$ is a parameter which permits us to deal with one particle Green function of the $2+1$ dimensional theory to
derive the effective action of the external fields as in Section \ref{topCS}. In fact, integrating out the spinor fields $\psi,\bar{\psi}$ in the related path integral yields the effective action
$$
S^\td_{eff}=G_{1D}(k_y) \int dx dt \zeta(x,t) \epsilon_{rs}\partial_r A_s .
$$
The coefficient $G_{1D}(k_y)$ is required to satisfy
\begin{equation}
\label{g1}
 \int G_{1D}(k_y) dk_y =N_1,
\end{equation}
where the first Chern number $N_1$ is given by (\ref{C1M}). Instead of the Cartesian coordinates we prefer to work
with the polar coordinates $k,\ \theta$, where 
$k_x=k\cos\theta ,\ k_y=k\sin\theta .$
Similar to (\ref{g1}) we would like to introduce  $G (\theta )$ satisfying
\begin{equation}
\label{gtd}
\int_0^{2\pi}  G (\theta ) d\theta =N_1,
\end{equation}
and define the $(1+1)$-dimensional  effective action as
\begin{equation}
\label{e2t}
S_{\td}=G (\theta )\int dx dt \zeta(x,t)\epsilon_{rs}\partial_r A_s .
\end{equation}
Although it can be deduced directly from  definition (\ref{gtd}), we can also obtain $G (\theta )$  by writing 
 the components of the     Abelian Berry gauge field (\ref{bgfi})  in polar coordinates:
$$
{\cal A}_{\theta} =-\frac{k}{2E(E+m)},\ {\cal A}_{k}=0.
$$
The Berry curvature remains the same
$$
{\cal F}_{k\theta}=\frac{1}{k}\left[\frac{\partial( k{\cal A}_{\theta}) }{\partial k}- \frac{\partial {\cal A}_{k} }{\partial \theta}\right]=-\frac{m}{2E^3},
$$
and allows us to  calculate  explicitly  $G (\theta  )$ as
\begin{eqnarray*}
G (\theta) &=& \frac{1}{2\pi}\int kdk{\cal F}_{k\theta} \\
       &=& -\frac{m}{4\pi}\int_\ssD \frac{dE}{E^2} = \frac{N_1}{2\pi}.
\end{eqnarray*}   
Now, one can define the one dimensional charge polarization\cite{kin-van, om} $P(\theta  )$ by
\begin{equation}
\label{defp}
\frac{\partial P(\theta) }{\partial \theta }\equiv G (\theta).
\end{equation}
Adopting the first Chern number calculated in (\ref{C1}), $N_1=1,$ we solve (\ref{defp}) by
\begin{equation}
\label{pit}
P(\theta)=\frac{\theta}{2\pi} .
\end{equation}
The physical observable is not directly the charge polarization given by $P(\theta)$ 
but the adiabatic change in $P(\theta)$ along a loop, which is  equal to 
$$
\Delta P=P(2\pi)-P(0)=1.
$$
The $(1+1)$-dimensional  action (\ref{e2t}) becomes
\begin{equation}
\label{e2t1}
S_{\td}=\frac{1}{2\pi} \int dx dt A_r\epsilon_{rs}\partial_s \zeta(x,t) ,
\end{equation}
for $N_1=1.$ Action (\ref{e2t1}) leads to 
the  current
$$
j_r= \frac{1}{2\pi} \epsilon_{rs}\partial_s \zeta(x,t),
$$
known as the Goldstone-Wilczek formula\cite{gw} and gives the charge
$$
Q=\frac{1}{2\pi}\int  \frac {\partial \zeta(x,t)}{\partial x}dx = \frac{1}{2\pi} \Delta \zeta .
$$
In fact, it corresponds to solitons on polyacetylene with charge $Q=1/2$
for  $\zeta$ changing from $0$ to $\pi$ and  $Q=1/3$ for $\Delta \zeta =2\pi /3 $
as it was obtained in \cite{gw}.

\setcounter{equation}{0}
\section{ $\bm{4+1}$ dimensional topological insulator and dimensional reduction to  $\bm{3+1}$ and $\bm{2+1}$ dimensions \label{foap}}

The topological field theory 
\begin{equation}
\label{cs5}
S^{\fid}_{eff}[A]=\frac{N_2}{24\pi^2}\int d^5x\epsilon^{ABCDE}A_A \partial_B A_C \partial_D A_E ,
\end{equation}
is designated as  the effective action  of the $4+1$ dimensional    TRI topological insulators in \cite{qhz}.
It follows from (\ref{CS}) by making use of relation (\ref{c2n}).
To derive the related second Chern number $N_2$ we deal with 
the $4+1$ dimensional    realization of the Dirac Hamiltonian (\ref{DH}) which is provided by 
\begin{align}
\label{alphabeta4+1}
\alpha_{1,2,3}=\left(
\begin{array}{cc}
  0 & i \sigma_{1,2,3} \\
  -i\sigma_{1,2,3} & 0 \\
\end{array}
\right),\ \ \
\alpha_{4}=\left(
\begin{array}{cc}
  0 & -1 \\
  -1 & 0 \\
\end{array}
\right),\ \ \
\beta=
\left(
\begin{array}{cc}
  1 & 0 \\
  0 & -1 \\
\end{array}
\right).
\end{align} 
Observing that 
$\rho_{i}= (i\sigma_1,i\sigma_2,i\sigma_3,-1),$ the non-Abelian Berry gauge fields can  be obtained from (\ref{BGFgeneral})  as
\begin{eqnarray}
  {\cal A}_1 =\frac{\sigma_{3}k_2-\sigma_{2}k_3-\sigma_{1}k_4}{2E (E+m)} ,\ \
  {\cal A}_2 =\frac{-\sigma_{3}k_1+\sigma_{1}k_3-\sigma_{2}k_4}{2E (E+m)} , \label{aa1}\\
  {\cal A}_3 =\frac{\sigma_{2}k_1-\sigma_{1}k_2-\sigma_{3}k_4}{2E (E+m)} ,\ \
  {\cal A}_4 =\frac{\sigma_{1}k_1+\sigma_{2}k_2+\sigma_{3}k_3}{2E (E+m)} . \label{aa2}
\end{eqnarray}
By definition the Berry gauge field corresponding to  the $4+1$ dimensional Dirac Hamiltonian  can also be derived by considering the explicit solutions
of the Dirac equation as it was done in \cite{rsfl}. They  work   in  the chiral representation, so that  the Berry gauge field components
which they obtain differ from (\ref{aa1}),(\ref{aa2}).

One can show that the field strength  components
${\cal F}_{ij} =\partial {\cal A}_j/ \partial k_i- \partial {\cal A}_i/ \partial k_j-i[{\cal A}_i,{\cal A}_j]$,
are 
\begin{eqnarray*}
  {\cal F}_{12} &=&\frac{1}{2E^{3}(E+m)}\left[\sigma_{3}(-E(E+m)+k_1^2+k_2^2)+\sigma_{2}(k_1k_4-k_2k_3)-\sigma_{1}(k_2k_4+k_1k_3)\right] ,\nonumber \\
  {\cal F}_{13}&=&\frac{1}{2E^{3}(E+m)}\left[\sigma_{2}(E(E+m)-k_1^2-k_3^2)+\sigma_{1}(k_1k_2-k_3k_4)+\sigma_{3}(k_1k_4+k_2k_3)\right] ,\nonumber \\
  {\cal F}_{14} &=&\frac{1}{2E^{3}(E+m)}\left[\sigma_{1}(E(E+m)-k_1^2-k_4^2)-\sigma_{2}(k_1k_2+k_3k_4)-\sigma_{3}(k_1k_3-k_2k_4)\right] ,\nonumber \\
  {\cal F}_{23} &=&\frac{1}{2E^{3}(E+m)}\left[\sigma_{1}(-E(E+m)+k_2^2+k_3^2)-\sigma_{2}(k_1k_2+k_3k_4)-\sigma_{3}(k_1k_3-k_2k_4)\right],\nonumber\\
  {\cal F}_{24} &=&\frac{1}{2E^{3}(E+m)}\left[\sigma_{2}(E(E+m)-k_2^2-k_4^2)-\sigma_{1}(k_1k_2-k_3k_4)-\sigma_{3}(k_1k_4+k_2k_3)\right],\nonumber\\
  {\cal F}_{34}&=&\frac{1}{2E^{3}(E+m)}\left[\sigma_{3}(E(E+m)-k_3^2-k_4^2)+\sigma_{2}(k_1k_4-k_2k_3)-\sigma_{1}(k_2k_4+k_1k_3)\right] \label{BC4+1} .
\end{eqnarray*}
Plugging them into (\ref{C2def}) and taking the trace yield
\begin{equation}
\label{C2_1}
N_2=\frac{3}{4\pi^2}\int (-\frac{m}{2E^5})d^{4}k.
\end{equation}
To calculate it explicitly,
we would like to deal with the four dimensional polar coordinates given by $k_1=k\cos\phi_1$, $k_2=k\sin \phi_1\cos\phi_2$, 
$k_3=k\sin\phi_1\sin\phi_2\cos\phi_3$ and $k_4=k\sin\phi_1\sin\phi_2\sin\phi_3,$ where the angles $\phi_1, \phi_2,\phi_3,$ respectively, take values in 
the intervals $[0,\pi],[0,\pi],[0,2\pi].$ 
The volume element  is $d^4k=k^3\sin^2\phi_1\sin\phi_2dk d\phi_1d\phi_2d\phi_3 .$
Hence, after the change of variable by (\ref{energy}), one can show that   (\ref{C2_1}) can be written as
\begin{equation}
\label{scnum}
N_2=\frac{3m}{4}\int_D\frac{m^2-E^2}{E^4}dE .
\end{equation}
When  $D$ is taken to be an  overlap of the $E>0$ and  $E<0$ domains, we may deal with  
\begin{equation}
\label{n21}
N_2=\frac{3m}{4}\int_{-\infty}^{m}\frac{m^2-E^2}{E^4}dE +\frac{3m}{4}\int_{-m}^{\infty}\frac{m^2-E^2}{E^4}dE=1.
\end{equation}

\subsection{Dimensional reduction to $\mathbf {3+1}$ dimensions \label{trbh4} }
Dimensional reduction  of the $4+1$ dimensional effective action given by (\ref{lag})  to 
$3+1$ dimensions can be described by the Lagrangian density 
\begin{equation}
\label{La31}
{\cal L}_\fd [\psi,\bar{\psi},A] =\bar{\psi} \left[\gamma^{\alpha}\left(p_\alpha+A_\alpha\right)+\gamma_4 \tilde{\theta} -m\right]\psi ,
\end{equation}
where $\alpha=0, \cdots ,3.$ The external field $\tilde{\theta}(x_\alpha)$  is the reminiscent of the gauge field $A_4.$
$\psi,\ \bar{\psi}$ fields can be integrated out as in Section \ref{topCS} through the one particle Green function of $4+1$ dimensional theory 
introducing the parameter $k_4$ by setting  $\tilde{\theta}=k_4+\theta (x_\alpha).$ By keeping track of the phase space volume one can obtain
the $3+1$ dimensional    effective action  as
\begin{equation}
\label{sefb}
S_{eff}^{\fd}=\frac{G_{3D}(k_4)}{4\pi}\int d^4x \theta\epsilon^{\alpha\beta\gamma\eta}\partial_{\alpha} A_{\beta} \partial_{\gamma} A_{\eta} ,
\end{equation}
where the coefficient is given through the condition 
\begin{equation}
\label{dg3}
\int G_{3D}(k_4) dk_4 =N_2 .
\end{equation} 
We would like to modify this construction by working  with the four dimensional polar coordinates  and  proposing that
 the  action describing the descendant theory is given by
$$
S_{\fd}=\frac{G_3(\phi_3)}{4\pi}\int d^4x \theta\epsilon^{\alpha\beta\gamma\eta}\partial_{\alpha} A_{\beta} \partial_{\gamma} A_{\eta} ,
$$
whose coefficient, like (\ref{dg3}), is required  to satisfy the  condition  
$$
\int_{0}^{2\pi} G_{3}(\phi_3)d\phi_3 =N_2 .
$$
Thus, the coefficient $G_{3}(\phi_3)$ can be obtained as
\begin{equation}
\label{g3n2}
G_{3}(\phi_3) =\frac{1}{32\pi^2 }\int \epsilon_{ijkl}\tr ({\cal F}_{ij}{\cal F}_{kl})k^3\sin^2\phi_1 \sin\phi_2 dkd\phi_1d\phi_2   =\frac{N_2}{2\pi},
\end{equation}
with  definition (\ref{scnum}) of the second Chern number $N_2.$

Similar to the one-dimensional  charge polarization (\ref{defp}) one can associate the coefficient $G_3(\phi_3)$ to $P_3 (\phi_3)$
 through the  relation\cite{qhz}
$$
\int_{0}^{2\pi}{\frac{\partial P_3 (\phi_3)}{\partial \phi_3}d\phi_3}\equiv\int_{0}^{2\pi} G_{3}(\phi_3)d\phi_3 = N_2.
$$
Hence the ``magnetoelectric polarization" can be obtained as
\begin{equation}
\label{P3D}
P_{3}(\phi_3)=\frac{N_2}{2\pi}\phi_3 .
\end{equation} 
Observe that like  the one-dimensional case,  for  $\Delta \phi_3=2\pi$ it changes by $\Delta P_{3}=1$ if we choose  $N_2=1$ as it is calculated in
(\ref{n21}). $P_{3}(\phi_3)$ depends linearly on $\phi_3$  due to the fact that the second Chern character corresponding to free Dirac particle 
depends only on  $k.$ 
Interacting  Dirac particles may give rise to polarizations which would not be linearly dependent on $\phi_3.$ 

By inserting definition (\ref{g3n2}) into (\ref{sefb}), the effective action  becomes
\begin {equation}
\label{ax}
 S_{\fd}=\frac{N_2}{8\pi^2}\int d^4x \theta \epsilon^{\alpha\beta\gamma\eta}\partial_\alpha A_\beta \partial_\gamma A_{\eta}.
\end{equation} 
It can be written equivalently  as
$$
 S_{\fd}=\frac{1}{4\pi}\int d^4x P_3(\theta) \epsilon^{\alpha\beta\gamma\eta}\partial_\alpha A_\beta \partial_\gamma A_{\eta},
$$
where $P_3(\theta )=N_2\theta /2\pi .$
This describes  the  axion electrodynamics which is invariant under the shift $\theta \rightarrow \theta+2\pi$ \cite{wil,vf}.
 
\subsection{A hypothetical model for $\mathbf {3+1}$ dimensional topological insulators \label{gkm}}

In spite of the fact that the underlying topological gauge theory (\ref{cs5}) is manifestly TRI, the 
theory given by the descendant
 action (\ref{ax}) is TRB except for the values $\theta=0,\pi .$ 
Nevertheless, we may deal with the TRB action  (\ref{ax}) but 
introduce a TRI hypothetical model generalizing the spin Hall effect for graphene\cite{km}.   
The current following from  action  (\ref{ax}) is 
$$
j^\alpha = \frac{N_2}{(2\pi)^2} \epsilon^{\alpha\beta\gamma\eta} \partial_\beta \theta \partial_\gamma A_\eta .
$$
Assuming $\theta=\theta(z)$ and  considering the in-plane electric field $ E_a(x,y) ;$ $a=1,2,$  we obtain  the current\cite{bdf}
\begin{equation}
\label{jd}
j_a= \frac{N_2}{(2\pi)^2}\partial_z\theta (z) \epsilon_{ab}E_b (x,y).
\end{equation}
The Hall current can  be introduced by   integrating (\ref{jd}) along the coordinate $z$ as
\begin{equation}
\label{dJ}
J_a (x,y)\equiv \int j_a dz =\sigma_{\ssH}\epsilon_{ab}E_b (x,y).
\end{equation}
It  leads to the surface Hall conductivity $\sigma_{\ssH}$ \cite{hm,qhz}
\begin{equation}
\label{S1h}
\sigma_\ssH=\frac{e^2}{\hbar} \frac{N_2}{(2\pi)^2}\int {\partial_z \theta dz}=\frac{e^2}{\hbar}\frac{N_2}{(2\pi)^2}\Delta \theta .
\end{equation}
Obviously we defined $\Delta \theta =\theta(\infty)- \theta(-\infty),$ which is non-vanishing for an adequate domain wall  or at an interface
plane between two samples.

Now we should define the second Chern number (\ref{scnum}) appropriately. 
We suppose that all negative energy states  are occupied till the Fermi level taken as the first positive energy value $E_\ssF =m,$ so that
we get
\begin{equation}
\label{niH}
N_2=\frac{3m}{4}\int_{- \infty }^{m}\frac{m^2-E^2}{E^4}dE =1/2.
\end{equation}
Considering  a plane of interface  which yields $\Delta \theta= 2\pi $ the Hall conductivity becomes
$$
\sigma_\ssH=\frac{e^2}{2h} .
$$

In representation (\ref{alphabeta4+1}), the $4+1$ dimensional Dirac Hamiltonian (\ref{DH}) is TRI where the time reversal operator can be taken as $T_\fid =\alpha_2\alpha_4K.$ However,  the Hamiltonian corresponding to  action (\ref{La31}) for $A_\alpha =0,$
\begin{equation}
\label{drh}
H_\fd= \bm \alpha\cdot \bm k+\alpha_4\tilde{\theta}+m \beta ,
\end{equation}
violates time reversal symmetry.
We will present a hypothetical model  which is time reversal invariant emulating the spin Hall effect for graphene.
Let $\alpha_i$ act on sublattices  with two Dirac points. Assume that
around these points 
which are interchanged under time reversal transformation, electrons are described by the  Hamiltonians as in (\ref{drh}).
Moreover, the third component of  the spin given by the Pauli matrix $s_z={\rm diag}(1,-1) $ is included and conserved.
Thus,  we propose to consider the Hamiltonian 
\begin{equation}
  \widetilde{H}_\fd = \widetilde{\bm \alpha}\cdot \bm k+\widetilde{\alpha}_4 \tilde{\theta} + \tau_z s_z \beta m,
 \label{H3t}
\end{equation}
where $\tau_z = {\rm diag } (1,-1)$ and in terms of $\alpha_{i}$ and $\beta$ given by (\ref{alphabeta4+1}) we defined
$$
\widetilde{\alpha}_i=(\alpha_{1},\tau_z\alpha_{2},\alpha_{3},\alpha_{4}). 
$$
Now, as in Section \ref{shg}, the time reversal operator interchanging the Dirac points and the third components of the spin can be defined   by $T=\tau_y s_yK,$ 
so that (\ref{H3t}) is TRI. Obviously, we can obtain   (\ref{H3t}) through the dimensional reduction from the $4+1$ dimensional action corresponding to the following free Hamiltonian  
\begin{equation}
  \widetilde{H} = \widetilde{\alpha}_i\cdot k_i + \tau_z s_z \beta m\equiv {\rm diag}(\widetilde{H}^{\uparrow +},\widetilde{H}^{\uparrow -},
  \widetilde{H}^{\downarrow +}, \widetilde{H}^{\downarrow -}).
 \label{H4t}
\end{equation}
As we show in  Appendix, the  four dimensional Hamiltonians defined by (\ref{H4t}) correspond to the second Chern numbers
$$
N_2^{\uparrow +}=N_2^{\uparrow -}=-N_2^{\downarrow +}=-N_2^{\downarrow -}=N_2,
$$
where $N_2$ is given by (\ref{scnum}).
Repeating the procedure yielding (\ref{dJ})-(\ref{niH}) in the presence of a domain 
wall we can obtain the dissipationless spin current as
$$
J_a^s=J_a^{\uparrow +}+J_a^{\uparrow -} -J_a^{\downarrow +}-J_a^{\downarrow -} =\sigma_{\ssSH}\epsilon_{ab}E_b(x,y),
$$
with the spin Hall conductivity 
$$
\sigma_\ssSH =\frac{e}{4\pi}\left( N_2^{\uparrow +}+N_2^{\uparrow -}-N_2^{\downarrow +}-N_2^{\downarrow -}\right)= \frac{e}{2\pi}.
$$
for $\Delta \theta= 2\pi$.
It is equal to the spin Hall conductivity for graphene (\ref{spcg}).

\subsection{Dimensional reduction to $\mathbf {2+1}$-dimensions}

The $2+1$ dimensional  Lagrangian density 
$$
{\cal L}_\thd [\psi,\bar{\psi},A] =\bar{\psi} \left[\gamma^\mu\left(p_\mu+A_\mu\right)+\gamma_3 \zeta_3 +\gamma_4 \zeta_4 -m\right]\psi,
$$
describes the dimensionally reduced theory. The fields 
$\zeta_3(x_\mu),\zeta_4(x_\mu)$ are the reminiscent of the gauge fields $A_3,A_4,$ of the $4+1$ dimensionally theory whose action is given by (\ref{lag})
for $d=4.$
By setting $\zeta_3(x_\mu)=k_3+\tilde{\phi}(x_\mu)$  and
$\zeta_4(x_\mu)=k_4+\tilde{\theta}(x_\mu),$ where $k_3, k_4$ are parameters playing the role of the momentum components in one particle Green functions, one can follow  the approach of Section \ref{topCS} to derive the
$2+1$ dimensional    effective action as
$$
S^{\thd}_{eff}=G_{2D}(k_3,k_4)\int d^3x \epsilon^{\mu\nu\rho} A_\mu \partial_\nu \tilde{\phi} \partial_\rho \tilde{\theta} .
$$
Its coefficient should fulfill the condition
\begin{equation}
\label{cfl}
\int G_{2D}(k_3,k_4) dk_3 dk_4 =N_2.
\end{equation} 
As in Section \ref{trbh4}, we consider the four dimensional polar coordinates and  propose that the action
\begin{equation}
\label{21a}
S_\thd =G_2(\phi_2,\phi_3)\int d^3x  \epsilon^{\mu\nu\rho} A_\mu \partial_\nu \tilde{\phi} \partial_\rho \tilde{\theta} ,
\end{equation}
describes the $2+1$ dimensional descendant theory.   
Obviously, like (\ref{cfl}) we pose the condition 
$$
 \int_0^{\pi}\int_0^{2\pi}d\phi_2d\phi_3 G_{2}(\phi_2,\phi_3) = N_2.
$$
This can be solved by
$$
G_2(\phi_2,\phi_3)= \frac{N_2}{4\pi}\sin\phi_2 .
$$
Proceeding as in \cite{qhz} we 
introduce the vector field
$$
\Omega_\mu \equiv \Omega_\theta \partial_\mu \theta +\Omega_\phi \partial_\mu \phi,
$$
however by adopting  the definitions
$$
\Omega_\theta= -\frac{N_2}{4}\cos \phi ,\ \Omega_\phi =-\frac{N_2}{4}\theta \sin \phi  .
$$
The field strength of the  field $\Omega_\mu$ is
$$
\partial_\mu\Omega_\nu-\partial_\nu\Omega_\mu =\frac{N_2}{2}\sin \phi \left(\partial_\nu \theta \partial_\mu \phi -\partial_\mu \theta \partial_\nu \phi \right).
$$
Let us define
$\phi =\phi_2 +\tilde{\phi}$ and $\theta =\phi_3 +\tilde{\theta}$ as slowly varying fields, so that at the first order in derivatives
we can write  $G_2(\phi,\theta)\partial_\mu \theta \partial_\nu \phi\approx G_2(\phi_2,\phi_3)\partial_\mu \tilde{\theta} \partial_\nu \tilde{\phi}.$
Therefore,  action (\ref{21a}) can be written in the form
\begin{equation}
\label{eOm}
S_{\thd}=\frac{1}{2\pi}\int d^3x \epsilon^{\mu\nu\rho} A_\mu \partial_\nu \Omega_\rho  .
\end{equation}
The current generated by the  field $\Omega_a,\ a=1,2,$ is 
$$
j_a^\Omega =\frac{e}{2\pi}\epsilon_{ab}E_b ,
$$
where the electric field  is given by  $E_a =\partial_a A_0- \partial_0 A_a.$
It can be interpreted as 
the spin current yielding the spin Hall conductivity $\sigma_\ssSH =e/2\pi .$
Hence,
by attributing the adequate time reversal transformation properties to the gauge field $\Omega_\mu ,$ action
 (\ref{eOm}) corresponds  to the TRI $2+1$ dimensional  model of \cite{km} which we discussed in Section \ref{shg},

Action (\ref{eOm}) generates the  electric current
$$
j^\mu=\frac{1}{2\pi} \epsilon^{\mu\nu\rho} \partial_\nu \Omega_\rho .
$$
For fields satisfying $\phi=\phi (x),\ \theta=\theta (y)$ it gives the total charge 
$$
Q=e \frac{N_2}{4\pi}\int dxdy \sin \phi \partial_x \phi \partial_y \theta = e \frac{N_2}{4\pi} \int_0^\pi\sin \phi  d\phi \int_0^{2\pi}d\theta =eN_2.
$$
On the other hand 
the three dimensional Skyrmion field $\bm n$ coupled to Dirac fermion in $2+1$ dimensions   yields the current\cite{jar}
$$
j^\ssT_\mu=\frac{1}{8\pi} \epsilon_{\mu\nu\rho} \bm n \cdot \partial^\nu \bm n \times \partial^\rho \bm n .
$$
The Skyrmion field configuration discussed in \cite{gs} satisfying  ${\bm n}^2=1$
possesses the charge
$
Q^\ssT =2e.
$
Hence, if we deal with $N_2=1,$ the Skyrmion theory can be  described 
for the field configurations satisfying 
$$
\sin \phi \left( \partial_\nu \theta \partial_\mu \phi -\partial_\mu \theta \partial_\nu \phi \right) =\frac{1}{4}  \bm n \cdot \partial_\nu \bm n \times \partial_\mu \bm n ,
$$
which leads to $j^\ssT_\mu =2j_\mu .$
In principle, this condition can be solved to obtain $\bm n$ in terms of the fields $\phi$ and $ \theta.$
 
Observe also that for the field configurations $\phi=\phi (t),\ \theta=\theta (y)$ the net charge flow in $x$ direction is
\begin{equation}
\int dtdyj_x =-N_2.
\label{ncf}
\end{equation}
Moreover, we can  introduce the magnetoelectric polarization in the form given (\ref{P3D}) by  defining it as 
$$
P_3 (\theta )=-\int_0^{\pi} d\phi \Omega_\phi/\pi =\frac{N_2}{2\pi} \theta .
$$
Then the pumped  charge (\ref{ncf}) can also be written as
$\Delta Q=\int  dP_3  =N_2$ which gives $\Delta Q=1/2$
for $N_2=1/2,$ as it is given for the Hall effect (\ref{niH}).

\section{Discussions}

The Foldy-Wouthuysen transformation  which diagonalizes the Dirac Hamiltonian is proven to be a powerful tool to
perform calculations in
 the effective field theory of the  $4+1$ dimensional TRI topological insulator.
The Foldy-Wouthuysen transformation is  employed to obtain the Berry gauge fields of Dirac Hamiltonians and through them we derived the first and second Chern characters explicitly. 
Then we demonstrated in a transparent manner that  the winding numbers of Dirac propagators are equal to the coefficients of the effective Chern-Simons actions  in $2+1$  as well as in $4+1$ dimensions. This construction  can be generalized to higher odd dimensions straightforwardly.   In the line of the graphene model \cite{km} we introduced a hypothetical model  leading to a dissipationless spin current in $3+1$-dimensions.
It can be helpful to understand some aspects of three dimensional TRI topological insulators  if we can show that it is somehow related to some realistic models.
Moreover, it seems that in terms of our explicit constructions one can  discuss $\mathbb{Z}_2$ topological classification of TRI insulators in a tractable fashion. 
In principle  our approach can be generalized to interacting Dirac particles where the related Foldy-Wouthuysen transformation  at least perturbatively exists.

\setcounter{equation}{0}

\setcounter{section}{1} \null

\renewcommand{\theequation}{\Alph{section}.\arabic{equation}}
\renewcommand{\thesection}{\Alph{section}}

\section*{Appendix }

The Hamiltonian (\ref{H4t}) which comprises $\tau_z$ and spin degrees of freedom  denoted  $\pm$ and $\uparrow \downarrow ,$ respectively, yields the $4+1$ dimensional    Dirac Hamiltonians 
\begin{align}
\widetilde{H}^{\uparrow+}
&=\alpha_{1}k_1+\alpha_{2}k_2+\alpha_{3}k_3+\alpha_{4}k_4+m\beta,
   &
\widetilde{H}^{\uparrow-}
&= \alpha_{1}k_1-\alpha_{2}k_2+\alpha_{3}k_3+\alpha_{4}k_4-m\beta,
 \label{H4up} 
   \\
\widetilde{H}^{\downarrow+}
&=\alpha_{1}k_1+\alpha_{2}k_2+\alpha_{3}k_3+\alpha_{4}k_4-m\beta,
    &
\widetilde{H}^{\downarrow-}
&=\alpha_{1}k_1-\alpha_{2}k_2+\alpha_{3}k_3+\alpha_{4}k_4+m\beta.
 \label{H4down}
 \end{align}
Let us first consider the two spin up Hamiltonians (\ref{H4up}). They yield slightly different 
non-Abelian Berry gauge fields 
\begin{eqnarray*}
  {\cal A}^{\uparrow\pm}_{1}=\frac{1}{2E(E+m)}(\pm\sigma_{3}k_2-\sigma_{2}k_3\mp\sigma_{1}k_4) ,\ 
  {\cal A}^{\uparrow\pm}_{2}=\frac{1}{2E(E+m)}(\mp\sigma_{3}k_1\pm\sigma_{1}k_3-\sigma_{2}k_4) ,\nonumber\\
  {\cal A}^{\uparrow\pm}_{3}=\frac{1}{2E(E+m)}(\sigma_{2}k_1\mp\sigma_{1}k_2\mp\sigma_{3}k_4) ,\ \ \ \
  {\cal A}^{\uparrow\pm}_{4}=\frac{1}{2E(E+m)}(\pm\sigma_{1}k_1+\sigma_{2}k_2\pm\sigma_{3}k_3) .
  \label{BGF4+1up}
\end{eqnarray*}
The corresponding field strengths can be calculated as
\begin{eqnarray*}
  {\cal F}_{12}^{\uparrow\pm} &=&\frac{1}{2E^{3}(E+m)}\left[\mp\sigma_{3}(E(E+m)-k_1^2-k_2^2)+\sigma_{2}(k_1k_4-k_2k_3)\mp\sigma_{1}(k_2k_4+k_1k_3)\right] ,\nonumber \\
  {\cal F}_{13}^{\uparrow\pm}
&=&\frac{1}{2E^{3}(E+m)}\left[\sigma_{2}(E(E+m)-k_1^2-k_3^2)\pm\sigma_{1}(k_1k_2-k_3k_4)\pm\sigma_{3}(k_1k_4+k_2k_3)\right] ,\nonumber \\
  {\cal F}_{14}^{\uparrow\pm} 
&=&\frac{1}{2E^{3}(E+m)}\left[\pm\sigma_{1}(E(E+m)-k_1^2-k_4^2)-\sigma_{2}(k_1k_2+k_3k_4)\mp\sigma_{3}(k_1k_3-k_2k_4)\right] ,\nonumber \\
  {\cal F}_{23}^{\uparrow\pm} &=&\frac{1}{2E^{3}(E+m)}\left[\mp\sigma_{1}(E(E+m)-k_2^2-k_3^2)-\sigma_{2}(k_1k_2+k_3k_4)\mp\sigma_{3}(k_1k_3-k_2k_4)\right],\nonumber\\
  {\cal F}_{24}^{\uparrow\pm} &=&\frac{1}{2E^{3}(E+m)}\left[\sigma_{2}(E(E+m)-k_2^2-k_4^2)\mp\sigma_{1}(k_1k_2-k_3k_4)\mp\sigma_{3}(k_1k_4+k_2k_3)\right],\nonumber\\
  {\cal F}_{34}^{\uparrow\pm}
&=&\frac{1}{2E^{3}(E+m)}\left[\pm\sigma_{3}(E(E+m)-k_3^2-k_4^2)+\sigma_{2}(k_1k_4-k_2k_3)\mp\sigma_{1}(k_2k_4+k_1k_3)\right] .\label{BC4+1up}
\end{eqnarray*}
Although they are different, they generate the same second Chern number equal to (\ref{C2_1}):
\begin{equation}
N_2^{\uparrow\pm}
 =\frac{1}{32\pi^2 } \int d^4k  \epsilon_{ijkl} \tr \left[ {\cal F}_{ij}^{\uparrow \pm}{\cal F}_{kl}^{\uparrow \pm}  \right] =\frac{3}{4\pi^2}\int (-\frac{m}{2E^5})d^{4}k.
\label{nya}
\end{equation}

The  non-Abelian Berry gauge fields corresponding 
to the two spin down  Hamiltonians (\ref{H4down}) can be shown to   satisfy  
$$
{\cal A}^{\downarrow\pm}_{i}(k_1,k_2,k_3,k_4)=(-1)^{\delta_{4i}}{\cal A}^{\uparrow\pm}_{i}(k_1,k_2,k_3,-k_4),
$$
without  summation over the repeated indices. Thus, the components of the related Berry curvature are
$$
{\cal F}_{ij}^{\downarrow\pm}(k_1,k_2,k_3,k_4) =(-1)^{\delta_{4i}+\delta_{4j}}{\cal F}_{ij}^{\uparrow\pm}(k_1,k_2,k_3,-k_4).
$$
They yield the  same second Chern number  which is given by (\ref{nya}) up to a minus sign: $N_2^{\downarrow \pm} =-N^{\uparrow\pm}_2.$  


\newcommand{\PRL}{Phys. Rev. Lett. }
\newcommand{\PRB}{Phys. Rev. B }
\newcommand{\PRD}{Phys. Rev. D }

\end{document}